\documentclass[final,authoryear,12pt]{elsarticle}
\usepackage{ifpdf}
\usepackage{graphicx,natbib,amssymb,lineno}
\makeatother
\begin{document}
\begin{frontmatter}


\title{The Effect of Cosmological Background Dynamics on the Spherical Collapse in MOND}

\author{M. Malekjani}

\address{Department of Physics, Faculty of Science, Bu-Ali Sina University, Hamedan
65178, Iran \\
malekjani@basu.ac.ir}

\author{H. Haghi}

\address{Department of Physics, Institute for Advanced Studies in Basic Science (IASBS), P.O.Box 45195-1159, Zanjan, Iran\\
haghi@iasbs.ac.ir }

\author{D. Mohammad-zadeh Jassur}

\address{Department of Theoretical Physics and Astrophysics, University of Tabriz, P.O. Box 51664, Tabriz, Iran \\
jassur@tabrizu.ac.ir}

\begin{abstract}
The effect of background dynamics of the universe on formation of
large scale structures in the framework of Modified Newtonian
Dynamics (MOND) is investigated. A spherical collapse model is used
for modeling the formation of the structures. This study is done in
two extreme cases: ({\it i}) assuming a universe with a low-density
baryonic matter without any cold dark matter and dark energy; ({\it
ii}) a dark energy dominated universe with baryonic matter, without
cold dark matter. We show that for the case ({\it ii}) the
structures virialize at lower redshifts with larger radii compared
to the low-density background universe. The dark energy slow downs
the collapse of the structures. We show that our results are
compatible with recent simulations of the structure formation in
MOND.
\end{abstract}

\end{frontmatter}

\section{Introduction}
Asymptotic flat rotation curves of spiral galaxies reveal the
missing mass problem in galactic scales \citep{bos81,rub85}. This
problem can be solved by cold dark matter (CDM) hypothesis, or
alternative theories of gravitation
\citep{car04,hall08,sob07,zhao07}. The CDM hypothesis as a dominant
paradigm together with the cosmological constant makes the standard
model of cosmology \citep{spe06}. Despite many efforts for direct
detection of dark matter, it has not been observed in
non-gravitational experiments. On the other hand, results from high
resolution N-body simulations do not seem to be compatible with the
observations on the galactic scales and also provide incompatible
spatial distribution of the sub-halos \citep{moo99,kly99,metz08}.\\
The Modified Newtonian Dynamics (MOND) paradigm is one of the
alternative theories that was proposed by Milgrom (1983) to explain
the dynamics of gravitational system. In MOND, the Newton's second
law is modified  as $\mu(a/a_{0}){\bf a} = {\bf a}_{N}+
\nabla\times{\bf H}$, where ${\bf a}_N$ is the Newtonian
acceleration vector, ${\bf a}$ is the MONDian acceleration vector,
$a=|\textbf{a}|$ is the absolute value of MONDian acceleration,
$\mu$ is an interpolating function which is used for transition from
the Newtonian to the MOND regime, and $a_{0}\simeq1.2\times10^{-10}$
ms$^{-2}$ is the MONDian threshold acceleration \citep{bek84}. The
value of the curl field ${\bf H}$ depends on the boundary conditions
and the spatial mass distribution, and vanishes only for some
special symmetries \citep{bek84}. This modification successfully
explains the flat rotation curves of spiral galaxies at the large
distances. Below a characteristic acceleration $a_0$, $\mu(x)\simeq
x$ and in the absence of curl field $H$, the Newton's second law is
approximated as $a=\sqrt{a_0a_N}$, the so called deep MOND regime.
MOND has recently been generalized to a general-relativistic version
which is called TeVeS \citep{bek04}.\\
One of the challenging issues for any gravitational theory is to
test its ability to reproduce the large scale structure formation in
the universe. In MOND, the large scale structure formation has been
studied by Sanders (1998) in a low-density universe, i.e. the
universe without dark energy and dark matter. He showed that in
MONDian cosmology, a patch of the universe smaller than the horizon
size evolves with a different rate than the background. Hence the
structures naturally can be formed through the scale dependent
dynamics. In this scenario small structures form faster than larger
ones (bottom-up hierarchical model). Note that in this model, every
point in the space can be assumed as a centner of collapse and the
initial fluctuation of baryonic matter has no significant role in
the formation of structures. In order to solve this problem, the
association of MOND force can be considered as a peculiar
acceleration in a finite-size region and the structures is formed
from the initial baryonic fluctuation \citep{san01}. It has been
shown that, this assumption leads to rapid growth of structures and
the power spectrum of the structure is similar to that of
obtained from $\Lambda$CDM universe \citep{san01}.\\
 In the scale of clusters of galaxy, MOND reduces the discrepancy between dynamical and detectable mass
 to a factor of 2-3 in the scales of clusters of galaxy, but does not remove it
 completely \citep{whi88,san03,pot05}. Sanders (2003) speculated on the
 remaining missing mass and showed that 2-eV neutrinos which can aggregate on the scales of clusters of galaxy
 are a possible candidate to interpret the remaining missing mass
 \citep{san03}.\\
Up to present day, several attempts in the cosmological structure
formation have done
by means of MONDian N-body simulation \citep{nus01,kneb04,linares}.\\
Malekjani, Rahvar and Haghi (2009, here after MRH09) showed that in
MOND, a uniform distribution of matter finally virializes with a
power law profile in which the virialization process takes place
gradually from the center of structure to the outer parts
\citep{malekjani09}. In this work, we discuss the effect of
\emph{background dynamics} on the growth rate of the structures. In
the case of universe dominated by cosmological constant, the
spherical collapse is similar to that in CDM model, except the
cosmological constant that changes the growth of the structure
though altering the background dynamics. The main gaol of the paper
is to understand the effect of background dynamics of the universe
on the evolution of spherical over-dense structures in the framework
of MOND. It is possible to achieve this goal even if we consider
only the pure baryonic matter as a single component of matter
content of the universe and avoid the complexity of neutrino problem
in the scales of clusters of galaxies. Using the simple baryonic
spherical collapse model in the context of MOND, we study the effect
of generic variable dark energy background model on the scenario of
structure formation.\\ The paper is organized as follows: In section
\ref{st_form} we introduce the spherical collapse in MOND. In
section \ref{bgrond_cosmo} we calculate the effect of background
dynamics on the structure formation in MOND. The paper is concluded
in section \ref{conclusion}.

\section{Spherical collapse in MOND}\label{st_form}
The following is a brief review of MONDian cosmology and the
spherical collapse in the framework of MOND \citep{san98}. In the
scales smaller than Hubble radius, the dynamics of the universe is
derived from the Hubble equation as
\begin{equation}
H^{2}(a)=H_{0}^{2}[\Omega_{b}^{0}a^{-3}+\Omega_{r}^{0}a^{-4}-(\Omega_{t}^{0}-1)a^{-2}+\Omega_{de}],
\label{hubb}
\end{equation}
where $H_0$ is the Hubble parameter at the present time,
$\Omega_b^0$, $\Omega_r^0$ and $\Omega_t^0$ are the baryonic matter,
radiation and total energy density parameters, respectively, and
$\Omega_{de}$ is the density parameter of dark energy. In MONDian
cosmology, one can consider a critical radius $r_c$ that inside it
the dynamics is MONDian (i.e., the acceleration is bellow $a_0$) and
outside it the dynamics is Newtonian. The critical radius and
corresponding mass scale are
\begin{equation}
\begin{array}{ll}
r_{c}=\frac{2a_0}{H_{0}^{2}\mid\Omega_{b}^{0}a^{-3}+2\Omega_{r}^{0}a^{-4}-2\Omega_{de}\mid},~~
M_c=\frac{a_0}{G}r_c^2.
\end{array}
\label{crit}
\end{equation}
The structures with mass $M>M_c$ are in the Newtonian regime, while
for $M<M_c$ are in the MONDian regime. At the beginning of
evolution, the dynamics of the structure is Newtonian and the
density contrast grows in proportion with scale factor. Afterwards,
the structure enters into the MONDian regime and the dynamics
significantly deviates from the background expansion (see Fig. 1 of
MRH09). In MONDian regime ( the scales $M<M_c$ ), the $\mu$ function
is approximated as $\mu(x)=x$ and the dynamics of structure is given
by
\begin{equation}
\ddot{r}=-\frac{\sqrt{GMa_{0}}}{r}, \label{monddyn}
\end{equation}
where $M$ is the mass and $r$ is the radius of structure.
Multiplying the both sides of Eq.(\ref{monddyn}) in $\dot{r}$ and
takeing the integration with respect to time, we obtain
\textbf{\begin{equation}
\dot{r}=\sqrt{v_i^2-2(GMa_0)^{\frac{1}{2}}\ln{\frac{r}{r_i}}},
\label{monddyn8}
\end{equation}}
where $r_i$ is the initial radius at the entrance time into the
MONDian regime and $v_i$ is the initial expansion velocity of the
structure\footnote{$v_i=H_ir_i(1-\delta_i)$ is the peculiar velocity
of the structure at the entrance time to MOND regime, where
$\delta_i$ is the initial density contrast at the entrance time.}.
Changing the time derivative to the derivative with respect to scale
factor $a$, Eq.(\ref{monddyn8}) is rewritten as
\begin{equation}\label{monddyn9}
\frac{dr}{da}=\frac{1}{aH(a)}\sqrt{v_i^2-2(GMa_0)^{\frac{1}{2}}\ln{\frac{r}{r_i}}},
\end{equation}
where $H(a)$ is given by Eq.(\ref{hubb}). The maximum radius of the
structure is $r_{m}=r_{i}e^{\alpha}$, where
$\alpha=v_{i}^2/\sqrt{4GMa_0}$ \citep{san98}. Throughout the
re-collapse, the global radial velocity converts to the dispersion
velocity and eventually the structure virializes
at $r_{vir}=r_{i}e^{\alpha-1/2}$ (for more details see MRH09).\\

\section{Background dynamics and the structure formation in MOND}
\label{bgrond_cosmo} Here, we calculate the influence of the
background dynamics on the MONDian structure formation for two
extreme cases of low-density universe and dark energy dominated
universe.

\subsection{Low-density background}
In the low-density background, there is no dark matter and dark
energy in the universe and the total density parameter is
$\Omega_{t}^{0}=\Omega_{b}^{0}+\Omega_r^0\sim 0.02$
\citep{wal91,car97}. We take three class of objects with the masses
of $10^{6}M_{\odot}$, $10^{11}M_{\odot}$ and $10^{13}M_{\odot}$
which represent the globular cluster, galaxy and cluster of
galaxies, respectively. By numerical integration of
Eq.(\ref{monddyn9}) the evolution of structures is obtained as a
function of redshift. The initial conditions ($r_i$, $v_i$), the
maximum radius, $r_m$, the virialization radius, $r_{vir}$, together
with corresponding redshifts ($z_m,z_{vir}$) are shown in Table
(\ref{jadval1}). The smaller structures enter the MOND regime and
virialize earlier than the larger ones. The dynamical behavior of
structure formation in MOND is similar to the hierarchical structure
formation in standard CDM model \citep{pady93,malek09}. Here we see
that the MONDian spherical collapse under a low-density background
obtains the virialization at high redshift and therefore can
interpret the rapid growth of density perturbations. This result is
also discussed in \citep{san01}. Sanders assumed two-field
Lagrangian-based theory of MOND and obtained the rapid growth of
density perturbation by solving the corresponding differential
equation in non-linear regime.

\subsection{Dark energy background}

An interesting theoretical candidate for dark energy is cosmological
constant with the time-independent equation of state $\omega=-1$
\footnote{The equation of state is $p=\omega \rho$}
\citep{zel67,wein89,car01}. However, the cosmological constant
provides excellent fit to the SNIa and CMB data, it suffers from the
coincidence and fine tuning problems at the early universe. One
solution to these problems is assuming the dynamical dark energy
models whose equation of state evolves with cosmic time. In these
models an evolving scalar field generates the energy and the
pressure of the dark energy, and provides a positive accelerating
universe. Here we use the variable dark energy model proposed by
Weetterich (2004) in which the equation of state is \citep{wet04}
\begin{equation}
\omega(a;b,\omega_{0})=\frac{\omega_{0}}{[1-b\ln(a)]^2},
\label{state}
\end{equation}
where $\omega_{0}$ is the state parameter at the present time, $a$
is the scale factor and $b$ is the binding parameter which is
related to the amount of dark energy in the universe. However,
$\omega$ evolves differently for various bending parameters, it
asymptotically converges to $\omega_0=-1$ at the present time (i.e.
$a=1$). In the variable dark energy dominated background, the
density parameter of dark energy in Eq.(\ref{hubb}) can be described
as
\begin{equation}
\Omega_{de}(a;b,\omega_{0})=\Omega_{de}^0
a^{-3[1+\overline{\omega}(a;b,\omega_{0})]},~~
\overline{\omega}(a;b,\omega_{0})=\frac{\omega_{0}}{[1-b\ln(a)]},
\end{equation}
where $\Omega_{de}^0= 8\pi G\rho^{0}_{de}/3H^{2}_{0}$  and $b=0$
corresponds to the standard cosmological constant model. We assume
no clustering for dark energy, but it can influence on the growth
rate of the background and consequently on the formation of
structures. The various density parameters for variable dark energy
model are

\begin{equation}
\Omega_{t}^{0}=\Omega_{b}^{0}+\Omega_{de}^{0}=1.0,
~~\Omega_{de}^{0}=0.98, ~~ \Omega_b=0.02, ~~ \Omega_{dm}^{0}=0.
\label{vari-d}
\end{equation}

\begin{figure}
\centerline{\includegraphics[width=0.8\textwidth]{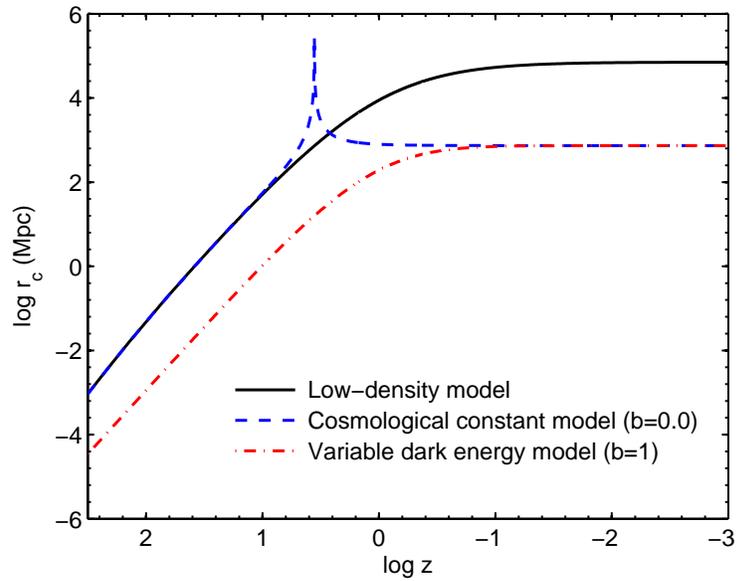}}
\caption{\small Log-log plot of the MONDian critical radius,
$r_{c}$, as a function of redshift for various cosmological
background models as described in legend. The cusp on the blue
dashed line is due to the equality epoch between baryonic matter and
dark energy, hence $r_c$ increases to infinity at this
time.}\label{r_c}
\end{figure}

\begin{figure}
\centerline{\includegraphics[width=.8\textwidth]{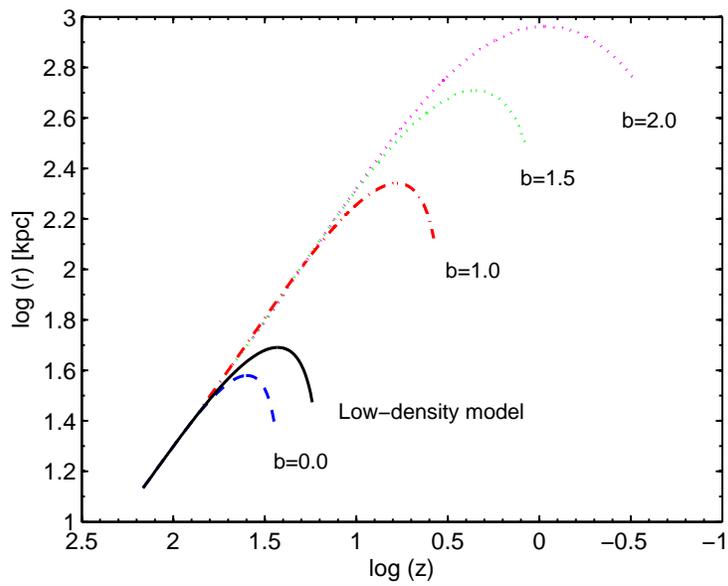}}
\caption{\small The evolution of a structure with the mass of
($10^{11}M_{\odot}$) from the entrance redshift, $z_{enter}$ until
the virialization stage for different values of bending parameter.
Increasing the bending parameter causes the slower collapse. The
case of $b=0$ refers to standard cosmological constant background.
For comparison, the evolution of the structure in the low-density
background is potted (black solid line).}\label{f11}
\end{figure}

\begin{figure}
\centerline{\includegraphics[width=.6\textwidth]{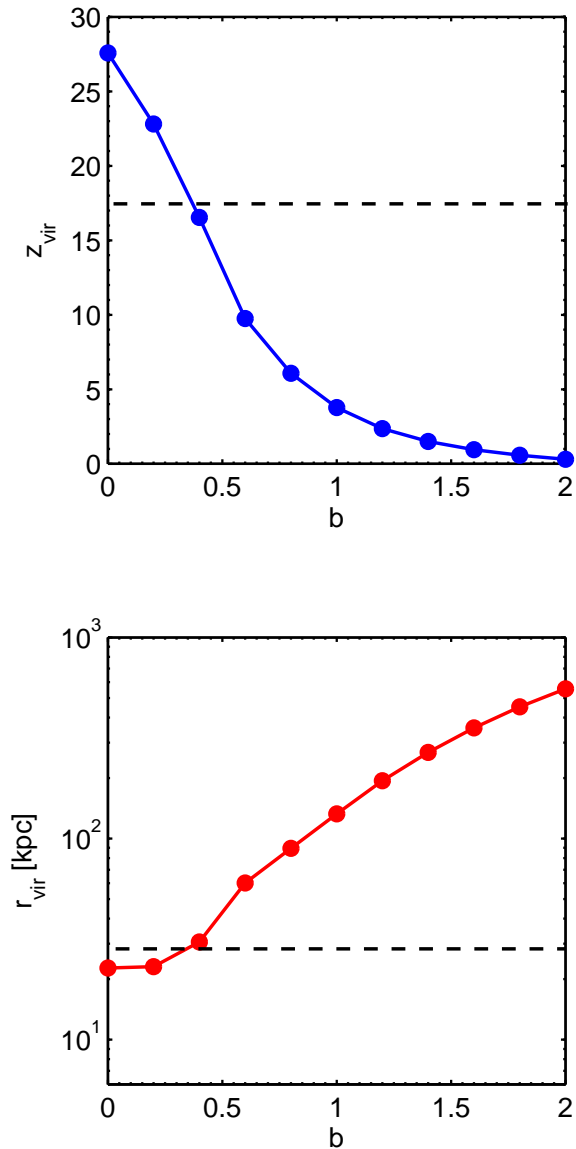}}
\caption{\small The dependency of the virialization redshift (upper
panel) and the virialization radius (lower panel) of
$10^{11}M_{\odot}$ structure as a function of bending parameter in
variable dark energy dominated background model. By increasing the
bending parameter the structure virializes at lower redshift with
larger radius. The horizontal dashed line indicates the
virialization in low-density background model. The virialization in
low-density model is equal with the virialization in variable dark
energy background with $b\sim 0.4$.}
 \label{f12}
\end{figure}

Figure (\ref{r_c}) shows the evolution of critical length scale
$r_c$, as a function of redshift $z$, for different cosmological
background models calculated from Eq. (\ref{crit}). In the case of
low-density model (black solid line), the critical radius increases
with the scale factor. It means that the MONDian domains gets larger
with time, and therefore the larger structures enter to the MONDian
regime after the smaller ones. The critical radius for the case of
standard cosmological constant model (blue-dashed line) evolves same
as the low-density model for $z\gg1$, since the contribution of
baryonic matter is dominated compare with cosmological constant in
the early times. The cusp in blue-dashed line indicants the equality
epoch of baryonic matter and dark energy, hence $r_c$ increases to
approach infinity at this time. After the equality epoch, the
cosmological constant is dominated and $r_c$ is smaller than that of
low-density model. This implies that in the universe dominated by
cosmological constant, the structure crosses $r_c$ and enters the
MOND regime later than low-density universe. In the case of variable
dark energy model with $b=1$ (red dotted-dashed line), $r_c$ is
smaller than other two models at any redshift. For example in the
case of $b=1$, the structure with the mass of $M=10^{11}M_{\odot}$
crosses the $r_c$ and enters the MOND regime at $z_{enter}=68$,
while for the low-density model, this happens at $z_{enter}=146$. It
should be noted that, because of the logarithmic term in Eq.
(\ref{state}), the critical radius for all models with
$b\neq0$ converges to the case of $b=0$ at $z\ll1$.\\
In order to obtain the evolution of the structure in MOND under the
variable dark energy background, we integrate the
Eq.(\ref{monddyn9}) by using Eq.(\ref{hubb}). For different values
of $b$, the entrance redshift, the initial radius, and the initial
expansion velocity at the entrance time for a galaxy with the mass
of $M=10^{11}M_{\odot}$ are shown in Table (\ref{jadval2}).
Increasing the bending parameter leads the structure to enter the
MOND regime at the lower redshift. Figure (\ref{f11}) shows the
evolution of galaxy mass structure for different cosmological
background models. The numerical results are summarized in the last
four columns of Table (\ref{jadval2}). The dependence of the
virialization redshift and the virialization radius of a galaxy mass
structure to the bending parameter are shown in Figure (\ref{f12}).
The structure virializes at lower redshift with larger radius for
the higher values of bending parameter. The virialization in
low-density model is equal to the virialization in variable dark
energy dominated background with $b\sim0.4$.

\subsection{Comparison with standard top-hat model}
In the framework of standard model of structure formation, the
baryonic structures are formed within the potential well of dark
matter. One of the simple analytical method for modeling the
structure formation  is the spherical collapse  with top-hat density
distribution model \citep{pady93,malek09}. In the standard top-hat
model, the virialization redshift of the galaxy mass structure takes
place at $z_{vir}\sim 0.5$ (see Tab.(1) of Ref.\citep{malek09}),
while in MOND, in the case of low-density background the
virialaization takes place at $z_{vir}\sim 17$. As we showed, the
variable dark energy background postpones the collapse of
structures. According to Table (\ref{jadval2}), the case of $b=2$
gives the virialization redshift that is compatible with standard
top-hat model.

\subsection{Comparison with MONDian N-body simulation}
\begin{figure}
\centerline{\includegraphics[width=.8\textwidth]{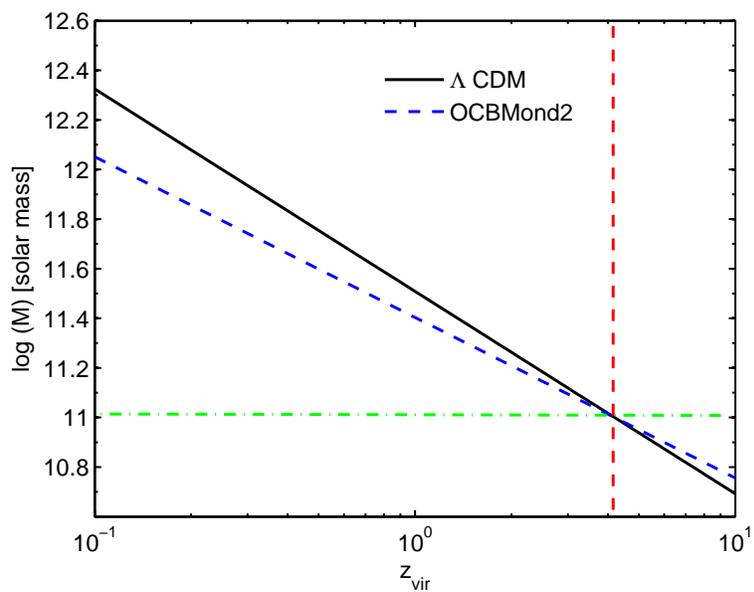}}
\caption{\small The  mass of structure in logarithmic scale as a
function of virialization redshift, based on Eq. (\ref{simulat}).
The horizontal dotted-dashed line represents the galaxy mass
structure, the vertical dashed line indicates the virialization
redshift of galaxy mass structure in OCBMond2, $\Lambda$CDM. The
virialization of a galaxy in $\Lambda$CDM and OCBMond2 is equal with
that of obtained in a variable dark energy model with $b=0.97$.}
 \label{f13}
\end{figure}

Recently, Linares et al. (2009) have presented cosmological N-body
simulation for the large scale structure evolution in the framework
of MOND. They studied the effects of the curl-field,
$\nabla\times{\bf H}$, upon the gravitational structure formation.
They also showed that the large scale structure formation is faster
in MOND due to strong gravity and found a correlation between the
mass of the structure and the formation redshift in $\Lambda$CDM and
MOND models. This correlation is given by
\begin{equation}
\begin{array}{lll}
\log(M/M_{\odot})=-0.8161\log{z_{vir}}+11.5086,~~ \Lambda CDM \\
 \log(M/M_{\odot})=-0.8981\log{z_{vir}}+11.4685,~~ OCBMond \\
\log(M/M_{\odot})=-0.6477\log{z_{vir}}+11.4036,~~  OCBMond2
\end{array}
 \label{simulat}
\end{equation}

where OCBMond denotes to the MOND model without considering
curl-filed in the solution of the MONDian Poisson's equation whereas
OCBMond2 is the solution with the influence of curl-field in the
process of structure formation. These linear relations are obtained
by the best-fit power laws to the scatter plot (see Fig. 11 of
Linares et al. 2009). According to Eq. (\ref{simulat}), the
evolution of structures in OCBMond2 is slightly faster than in
$\Lambda CDM$, hence the high-mass objects appear to form at earlier
times. There is also a significant difference between OCBMond and
OCBMond2 with the former being closer to $\Lambda CDM$. This
underlines the speed-up impact of the curl-field for the structure
formation which is similar to the effect of dark energy. In Figure
(\ref{f13}), we plot the mass of structure as a function of
virialization redshift, based on Eq.(\ref{simulat}). Here, we
compare the results of $\Lambda$CDM and OCBMond2 simulations with
the results of variable dark energy background model. Since the curl
field $H$ is neglected in OCBMond simulation and this model can not
justified in a fully three dimensional N-body simulation, therefore
we do not consider OCBMond simulation in this comparison. In the
case of OCBMond2 and $\Lambda$CDM the virialization redshift of
galaxy mass structure is $z_{vir}\sim 4$ which is comparable with
the virialization of galaxy in a variable dark energy background
model with $b=0.97$ (see Fig.\ref{f13}). Here we assume that the
mass of structure does not vary with time after the virialization.
Therefore today's mass of the structure is equal with the mass of
structure at the virialization time. This comparison can be
considered as a technical tool for studying the background fluid of
the universe. For example, here we see that the bending parameter
$b$ of the dark energy fluid is constrained as $b=0.97$, using the
spherical collapse in MOND.

\section{Conclusion}\label{conclusion}
In this paper we dealt with the problem of the structure formation
in the framework of Modified Newtonian Dynamics (MOND) under the
influence of variable dark energy dominated background model. We
utilized the spherical collapse model and showed that the
virialization of structures depends strongly on the background
cosmology. We showed that the variable dark energy dominated
background postpones the virialization of structures. Here the
results of three background models including the low-density model,
the standard cosmological constant model ($b=0$), and the variable
dark energy model ($b\neq0$) were compared. In the cosmological
constant model, the structures virialize faster than other models
(i.e., in the higher redshift). For variable dark energy model,
increasing the bending parameter $b$ causes the structure viralizes
at lower redshift with larger radius. Therefore the variable dark
energy model puts of the spherical collapse to the later times. The
case of low-density model has an intermediate behavior such that the
virialization redshift in this model corresponds with $b=0.4$ in
variable dark energy model. Finally, we compared the virialization
of structures under the variable dark energy model with the recent
results of MONDian N-body simulations. We showed that the various
models of simulation are consistent with the
variable dark energy model with different bending parameter.\\
It should be noted that most of the galaxies are formed within the
host environment such as cluster of galaxies and influenced by
external gravity. In our model, the structures assumed to be
isolated system which is acceptable approximation when the external
acceleration field is lower than the critical acceleration. Due to
nonlinearity of Poisson's equation in MOND, the strong equivalence
principle is violated and consequently the internal dynamics is
affected by the external filed. Therefore MOND predicts different
evolution for structures within the external gravity background. For
studying the background dynamics, we assumed the simple baryonic
spherical collapse model by avoiding the complexity of neutrino
problem which can be appeared in the scales of clusters of
galaxies.\\

\textbf{ Acknowledgment}\\
 We thank the anonymous referee for
useful comments. We also thank A. Moradi for reading the manuscript
and giving useful comments.

{}

\newpage
\begin{table}
\begin{center}
\caption{ Numerical results for the evolution of the structure in
the low-density background model for various mass scales in MOND.
First column indicates the mass of the structure. Second column is
the redshift of entrance time to the MOND regime. The third column
is the initial radius of the structure at the entrance time. The
forth column is the initial density contrast at the entrance time.
Fifth column denotes the initial velocity at the entrance time.
$r_m$ and $z_m$ are the maximum radius and corresponding redshift.
$z_{vir}$ and $r_{vir}$ are the virialization redshift and the
virialization radius of the structure.\label{jadval1}}
\begin{tabular}{|c|c|c|c|c|c|c|c|c|}
 \hline
 $M$               & $z_{enter}$ & $r_{i} [kpc]$ & $\delta_i\times 10^{-5}$ &$v_{i} [km/s]$  &$z_m$ & $r_m [kpc]$   & $z_{vir}$& $r_{vir} [kpc]$\\
 \hline
 $10^{6}M_{\odot}$ & $848$       & $0.033$       & $1.30$ &$14.5$          &$304$ & $0.079$       & $200$     & $0.047 $ \\
 \hline
 $10^{11}M_{\odot}$& $147$       & $13.4$        & $7.48$ &$315 $          &$27.5$ &$48$          & $17.5$   &$28.4$\\
 \hline
 $10^{13}M_{\odot}$&$74.5$       & $123.7$       & $14.60$ &$1097$         &$8.4$ &$591$     &$4.8$      &$339$\\
\hline
\end{tabular}
\end{center}
\end{table}

\begin{table}
\begin{center}
\caption{Numerical results for the evolution of galaxy mass scale
($10^{11}M_{\odot}$) with the variable dark energy background in
MOND. First column indicates the bending parameter of variable dark
energy. The second column is the entrance redshift to the MONDian
regime. Third column is the initial radius of the structure at the
entrance time. Fourth column is the initial velocity of the
structure at the entrance time. $r_m$ and $z_m$ are the maximum
radius and corresponding redshift. $z_{vir}$ and $r_{vir}$ are the
virialization redshift and the virialization radius of the
structure.\label{jadval2}}
\begin{tabular}{|c|c|c|c|c|c|c|c|}
\hline
  $b$ & $z_{enter}$ & $r_i [kpc]$ & $v_i [km/s]$ & $z_m$& $r_m [kpc]$ &$z_{vir}$ & $r_{vir} [kpc]$\\
    \hline

 $0.0$& $146$ & $13.4$ & $285$& $40$& $37.5$ & $27$ & $22.7$ \\
  \hline
$1.0$& $68$ & $29.0$ & $401$ & $6$ & $219.2$ &$4$ & $132.9$\\
  \hline
$1.5$& $45$ & $43.8$ & $442$ & $2.2$ & $510.8$ &$1.2$ & $309.8$\\
  \hline
$2.0$& $36$ & $53.7$ & $475$ & $1.0$ & $916.1$ &$0.3$ & $555.6$\\
\hline
\end{tabular}
\end{center}
\end{table}

\newpage

\begin{thebibliography}{00}
\expandafter\ifx\csname
natexlab\endcsname\relax\def\natexlab#1{#1}\fi
\expandafter\ifx\csname url\endcsname\relax
  \def\url#1{\texttt{#1}}\fi
\expandafter\ifx\csname urlprefix\endcsname\relax\def\urlprefix{URL}\fi


\bibitem[{Bekenestein (2004)}]{bek04} Bekenestein, J. D., 2004. Phys.
Rev. D. 70, 3509.
\bibitem[{Bekenestein \& Milgrom (1984)}]{bek84}
Bekenstein, J. D., Milgrom M., 1984. ApJ 286, 7.
%
\bibitem [{Bosma (1981)}]{bos81} Bosma, A., 1981. AJ, 86, 1825.
%
\bibitem[{Carlberg et al., (1997)}]{car97}
 Carlberg, R.G., et al. 1997, preprint, astro-ph/9706115.

\bibitem[{Carroll (2001)}]{car01}
 Carroll, S. M., 2001.  "The Cosmological Constant", Living Rev. Relativity 4, 1.

\bibitem[{Caroll et al. (2004)}]{car04}
 Caroll, S. M., {\it et al.}, 2004. Phys. Rev. D 70, 043528.

\bibitem[{Halle et al. (2008)}]{hall08}
  Halle, A., Zhao, H. S., Li, B., 2008. ApJS 177, 1.

\bibitem[{Klypin et al. (1999)}]{kly99}
Klypin, A. {\it et al.}, 1999. ApJ 522, 82.

\bibitem[{Knebe et al. (2004)}]{kneb04}
Knebe A., {\it et al.}, 2004. ApJ 603, 7.

\bibitem[{Llinares et al. (2008)}]{linares}
 Llinares C., Knebe A., Zhao H., 2008. MNRAS 391, 1778.

 \bibitem[{Malekjani et al. (2009a)}]{malekjani09}
 Malekjani, M., Rahvar, S., Haghi, H., 2009. ApJ 694, 1220, (MRH09).

\bibitem[{Malekjani et al. (2009b)}]{malek09}
 Malekjani, M., Rahvar, S., Jassur, D. M. Z., 2009. New Astronomy 14,
 398.

\bibitem[{Metz et al. (2008)}]{metz08}
 Metz M., Kroupa P. \& Libeskind N., 2008. ApJ 680, 287.

\bibitem[{Milgrom (1983)}]{mil83}
 Milgrom, M., 1983. ApJ 270, 365.

\bibitem[{Moore et al. (1999)}]{moo99}
 Moore, B. {\it et al.}, 1999. ApJ 524, L19.

\bibitem[{Nusser (2002)}]{nus01}
 Nusser,  A., 2002. MNRAS 331, 909.

\bibitem[{Padmanabhan (1993)}]{pady93}
Padmanabhan, T., 1993. Structure Formation in the Universe.
Cambridge University Press.

\bibitem[{Pointecouteau \& Silk (2005)}]{pot05}
Pointecouteau, E., \& Silk, J. 2005, MNRAS, 364, 654.

\bibitem[{Sanders (1998)}]{san98}
 Sanders, R. H., 1998. MNRAS 296, 1009.

\bibitem[{Sanders (2001)}]{san01}
Sanders, R. H., 2001. ApJ 560, 1.

\bibitem[{Sanders (2003)}]{san03}
Sanders 2003, MNRAS, 342, 901 whi88,san03.

\bibitem[{Sobouti et al. (2009)}]{sob07}
 Sobouti, Y., Hasani Zonoozi, A., Haghi, H., 2009. A \& A 507, 635.

\bibitem[{Spergel et al. (2006)}]{spe06}
Spergel, D. N. {\it et al.}, 2007. ApJS 170, 377.

\bibitem[{The \& White (1988)}]{whi88}
The \& White 1988, AJ, 95, 1642.

\bibitem[{van Albada et al. (1985)}]{rub85}
van Albada, T. S. {\it et al.}, 1985. ApJ 295, 305.

\bibitem[{Walker et al. (1991)}]{wal91}
 Walker, T.P., Steigman, G., Schramm, D.N., Olive, K.A., Kang, H., 1991. ApJ. 376,
 51.

\bibitem[{Weinber (1989)}]{wein89}
 Weinberg, S., 1989.  Rev. Mod. Phys. 61, 1.

\bibitem[{Wetterich (2004)}]{wet04}
  Wetterich C. 2004. Phys. Lett. B 594, 17.

\bibitem[{Zeldovich et al. (1967)}]{zel67}
 Zeldovich, Ya. B., Pis'ma Zh. Eksp. Teor. Fiz., 6, 883 (1967).

\bibitem[{Zhao (2007)}]{zhao07}
 Zhao, H. S., 2007. ApJ 671, L1.

\end{thebibliography}
\end{document}